\documentclass[12pt]{article}
\usepackage{graphicx}
\usepackage{color}
\usepackage{amsmath,slashed, amssymb}
\usepackage{fancyhdr}
\usepackage[titletoc]{appendix}
\numberwithin{equation}{section} 

\def\be{\begin{equation}}
\def\ee{\end{equation}}
\def\beq{\begin{eqnarray}}
\def\eeq{\end{eqnarray}}
\def\bea{\begin{eqnarray*}}
\def\eea{\end{eqnarray*}}

\def\centeron#1#2{{\setbox0=\hbox{#1}\setbox1=\hbox{#2}\ifdim
\wd1\rangle\wd0\kern.5\wd1\kern-.5\wd0\fi
\copy0\kern-.5\wd0\kern-.5\wd1\copy1\ifdim\wd0\rangle\wd1
\kern.5\wd0\kern-.5\wd1\fi}}
\def\ltap{\;\centeron{\raise.35ex\hbox{$\langle$}}{\lower.65ex\hbox{$\sim$}}\;}
\def\gtap{\;\centeron{\raise.35ex\hbox{$\rangle$}}{\lower.65ex\hbox{$\sim$}}\;}

\newcommand{\newc}{\newcommand}
\newc{\qbar}{{\overline q}}
\newc{\Kahler}{K\"ahler }
\newc{\deltaGS}{\delta_{\rm GS}}

\begin{document}
\begin{titlepage}
\begin{flushright}
{
SCIPP 18/01
}
\end{flushright}

\vskip 1.2cm

\begin{center}

{\LARGE\bf Some Remarks on Anthropic Approaches to the Strong CP Problem}

\vskip 1.4cm

{\large Michael Dine$^{(a)}$, Laurel Stephenson Haskins$^{(b)}$,\\
 Lorenzo Ubaldi$^{(c)}$, and Di Xu$^{(a)}$ }
\\
\vskip 0.4cm
{\it $^{(a)}$Santa Cruz Institute for Particle Physics and
\\ Department of Physics, University of California at Santa Cruz \\
     Santa Cruz CA 95064  } \\
     ~\\
{\it $^{(b)}$Racah Institute of Physics, \\
Hebrew University, Jerusalem 91904, Israel 
} \\
~\\
{\it $^{(c)}$ SISSA International School for Advanced Studies, \\
Via Bonomea 265, 34136 Trieste, Italy
}
\vspace{0.3cm}

\end{center}

\vskip 4pt

\begin{abstract}
The peculiar value of $\theta$ is a challenge to the notion of an anthropic landscape.  We briefly review the possibility
that a suitable axion might arise from an anthropic requirement of dark matter.
We then consider an alternative suggestion of Kaloper and Terning that $\theta$ might be
correlated with the cosmological constant.
We note  that in a landscape one expects that $\theta$ is determined by the expectation value of one or more axions.  We discuss
how a discretuum of values of $\theta$ might arise with an energy distribution
dominated by QCD, and find the requirements
to be quite stringent.  Given such a discretuum, we find
no circumstances where  small $\theta$ might be
selected by anthropic requirements on the cosmological constant.

\end{abstract}

\end{titlepage}

\section{Introduction}

At present, however frustrating it may be, the most compelling solution we have to the cosmological constant (c.c.) problem is provided by the 
anthropic landscape\cite{weinbergcc,bp,susskindlandscape,kklt}.  The picture scored an enormous success with the {\it discovery}
of the dark energy\cite{cc1,cc2,cc3}, with a value only somewhat smaller than expected from the simplest version
of Weinberg's argument.  More refined versions of the argument may come closer. This success has raised the specter that anthropic considerations may play an important
role in determining {\it all} of the laws of physics.  But there are good reasons for skepticism.  One might expect, in such a picture, that the parameters of the Standard Model should either be anthropically determined or should be random numbers.  A principled objection to these ideas, then, emerges from the fact that
some parameters of the Standard Model appear to be neither random nor anthropically constrained\cite{banksdinegorbatov}.
Possibly the most dramatic of these is the $\theta$ parameter\cite{banksdinegorbatov,donoghue, Ubaldi:2008nf}.  If a landscape picture is
ultimately to make sense, one needs to find correlations between the $\theta$ parameter 
and other quantities that are anthropically constrained.  For example, it is conceivable that
dark matter is an anthropic requirement\cite{aguirretegmark}, and it might be that the most efficient way to obtain dark matter in a landscape is through
an axion.  This is a tall order.  In particular, the requirement of a dark matter axion does not necessarily imply a Peccei-Quinn symmetry
of sufficient quality to explain $\theta < 10^{-10}$\cite{cdf}.   

The notion of a landscape can hardly be considered well-developed; we don't have gravity theories in which we can reliably demonstrate the existence of even
a small number of non-supersymmetric vacua, whereas we require a vast number of states. 
At best, we have toy models for the phenomenon, with which we can develop speculations
about questions like statistical distributions of Standard Model parameters.  Because we require both
an understanding of the microphysics of the landscape and its cosmology, even the precise questions of greatest interest are not clear.  But we might like to know, among the subset of states with Standard Model degrees of freedom at low energies,
what is the distribution of Lagrangian parameters, $P(\Lambda,m_H^2, g_i,y_{f \bar f},\theta)$, where $\Lambda$ is the c.c., $g_i$ are the gauge couplings, $y_{f\bar f}$ are the Yukawa couplings, $m_H^2$ is the Higgs mass, and $\theta$ is the QCD $\theta$ parameter.
It seems plausible, as Weinberg assumed, that the cosmological constant is a uniformly distributed random variable near its
observed value.  The question of {\it naturalness} of the Higgs mass is the question of whether the same is true
for the Higgs mass.  If the distribution is uniform, then one would seem to require an anthropic explanation of the Higgs mass, or alternatively rely on an extraordinary piece of luck.   On the other hand, supersymmetry or dynamics could enhance the probability that one finds the mass near zero, providing a realization of naturalness in a landscape framework.

The problem of $\theta$ in a landscape is that, absent a light axion, it is not clear why a distribution of $\theta$ should peak at $\theta=0$.  As we will explain below, in a model, say, like that of KKLT\cite{kklt}, one would expect that $\theta$ is a discrete, uniformly distributed random variable.  As we have noted, there is no obvious anthropic preference for extremely small $\theta$.  For critics of the landscape program,
this is perhaps the most principled argument that the landscape idea may not be correct.  For it to survive, one almost certainly
has to find correlations between $\theta$ and other quantities which {\it are} anthropically constrained.

It is possible that small $\theta$ is selected by some other consideration, dark matter being
one possible candidate. Recently,
Kaloper and Terning (KT)\cite{kt} have put forward another proposal to account for small $\theta$, which would correlate the value of $\theta$ with the problem of the 
cosmological constant.  In this note, we will attempt  to flesh out this proposal, determining what is required at a microscopic
level to realize their picture, and what might be the parameters of such a model and their possible
distributions.   Then we ask in what range of parameters one would in fact account for a small value of $\theta$.

KT assume that the cosmological constant has two contributions, one from a structure
similar to that of Bousso and Polchinski (BP)\cite{bp}, which we will denote by $\Lambda_{BP}$, and another, independent, one from QCD.
\beq \label{cc1}
\Lambda = \Lambda_{BP} + \frac{1 }{ 2} m_\pi^2 f_\pi^2 \theta^2.
\eeq
The typical spacing of values of $\Lambda_{BP}$ will be denoted as $\Delta \Lambda$; one can think of this as roughly some fundamental scale raised
to the fourth power  (say
$M_p^4$) divided by the number of states.
KT assume that $\theta$ is a continuous variable, and argue that, for a range of $\Delta \Lambda$, a small, negative value of $\Lambda_{BP}$ will be compensated by a small $\theta$, bringing the c.c.~into the anthropically allowed range. 

But this picture raises several puzzles.
In our generic landscape picture, above, we would expect that $\theta$ is some combination of axion expectation values in the underlying theory.   If $\theta$ can be considered to be independent of the values of the fluxes, and if QCD is the dominant source of the $\theta$ potential,
then we would seem to have a conventional axion.

On the other hand, we might expect the axion expectation values to be determined, in a scenario like that of BP, by values of fluxes
or other quantities which label the different states.  In this picture, the $\theta$ angles would take on random, discrete values,
not obviously correlated with other quantities.  The KKLT\cite{kklt} model provides a sharp implementation of this
picture.   In that model, the axion is not light. There, $\theta$ is fixed by the expectation values of \Kahler moduli, which are themselves fixed
by the (random) expectation value of the superpotential of complex structure moduli.  This superpotential, $\langle W \rangle$, itself is complex.
  In order that $\theta$ vanish, one might require that the expectation value of the
\Kahler modulus be real.  This would be the case among the tiny subset of fluxes (a fraction of order $(1/2)^N$, where $N$ is the number
of flux types) which conserve CP\cite{dinesun}.  But this might not
be enough once one has identified the origin of the CKM phase.  One would require an additional layer of structure, similar 
to that of Nelson-Barr models\cite{nelson,barr}.

It is also possible that in the would-be string landscape there are typically one or more light axions, yielding a conventional Peccei-Quinn solution of the
strong CP problem.  Such a prospect leads to the notion of an {\it axiverse}\cite{axiverse}.  The criteria for a solution
of the strong CP problem in such a setting have
been elaborated in \cite{moduliinflation}.  

All of this is to say that a ``conventional" landscape picture suggests that $\theta$ should 
 be a discrete variable, with no
obvious correlation with questions like the value of the c.c..

An alternative is that the axions are heavy, with, for a fixed choice of the fluxes, a {\it very} large number of nearly degenerate minima, with the degeneracy lifted
by (the {\it very} small) effects of QCD.  For $\theta$ to play any role in determining the c.c., one needs, in this case, very small steps in $\theta$, $\Delta \theta$.  We call
$\Lambda_a$ the upper limit of the anthropic window, which we assume close to the observed value,
i.e.~$\Lambda_a \sim 10^{-47}$ GeV$^4$. Then a minimal requirement on $\Delta \theta$ is that it leads to steps in
$V$, $\Delta V$, small enough to bracket $\Lambda_a$:
\beq \label{Vstep}
\vert \Delta V \vert \approx \frac{1}{2} m_\pi^2 f_\pi^2  \Delta \theta^2 < \Lambda_a
\eeq
yielding
\beq
\Delta \theta < 10^{-22}.
\eeq

In this case, it would not be critical that fluxes conserve CP.  The minimum of the $\theta$ potential, for any
choice of fluxes, would
lie at the CP conserving point for QCD; smaller $\theta$ contributions to the c.c.~would be associated with
smaller values of $\theta$.  We will develop several models of this type, in the spirit of outlining some of the ingredients
required to achieve a correlation between $\theta$ and the c.c.  Modest numbers of states can be accounted for within
conventional field theory principles.  The extension of these models to account for small enough $\Delta \theta$ requires
features which are not particularly plausible.
 Conceivably there is some more plausible structure which could give
rise to these features.

Allowing for such a structure, we then ask:  does the requirement that the observed cosmological constant lie within the anthropic window, $0 < \Lambda < \Lambda_a$, favor any particular range of $\theta$?   We will survey a two-parameter
space: $\Delta \Lambda$, the typical  spacing of cosmological constants in the flux vacua; and $\Delta \theta$.  
We find that throughout this parameter space, the anthropic constraint on the dark energy favors {\it large} $\theta$.

The rest of this paper is organized as follows.
In section \ref{anthropicaxions}, we review the problem of obtaining suitable axions assuming an anthropic requirement of dark matter\cite{cdf}.  We consider 
models where a discrete $Z_N$ symmetry accounts for an accidental Peccei-Quinn symmetry, and ask whether minimal requirements for dark matter
yield an $N$ large enough to account for $\theta < 10^{-10}$.  We also mention the (possibly different) expectations for string theory.

In section \ref{modelswhichwork}, we consider models which yield a discretuum for  
$\theta$.   We first note that the {\it irrational axion}\cite{irrationalaxion} has many of the desired
features.  It is not clear that such a structure ever arises in some more fundamental theory (string theory), and we will
see that, in any case,  it would tend to predict large $\theta$.  We then construct two models which implement at least some aspects of the KT program.  One involves a single axion coupled to a {\it very} large additional gauge group; the other involves multiple
axions and requires an intricate discrete symmetry.
These models are useful in that they {\it do} allow us to address a subset of the questions we have raised:
\begin{enumerate}
\item  In the theory $\theta$ is discrete, and the potential behaves as $V = - m_\pi^2 f_\pi^2\cos \theta$.
\item  The system is described in terms of two parameters, $\Delta \Lambda$, the typical  spacing of the BP contribution to the c.c.\footnote{We will use ``BP" to refer more generally to features of the theory, other than $\theta$, which allow for many possible values of the c.c.}, and 
$\Delta \theta$, the spacings in $\theta$.
\end{enumerate}
We will describe another model in the appendix.  In section~\ref{canceling}, we will ask, in terms of these parameters, where are the bulk of the states which satisfy the anthropic condition.  We will see that throughout the parameter          
space, cancellation of the c.c.~(to within anthropic constraints) is most effective at $\theta \sim 1$.

\section{Anthropic Axions}
\label{anthropicaxions}

It is conceivable that dark matter, with something close to its observed density, is an anthropic requirement\cite{aguirretegmark}.  In that case, the question becomes:
does the requirement of dark matter lead to a Peccei-Quinn symmetry of high enough quality to account for the smallness of $\theta$\cite{cdf}?
 One might imagine that in a landscape
setting, an axion might be a favorable dark matter candidate.  Models based on string constructions, for example, often have many axions, and there may be a significant fraction of the space of vacua in which one or more of these is very light.  We might model this by a field, $\phi$, subject to a $Z_N$ symmetry,
\beq
\phi \rightarrow e^{\frac{2 \pi i }{ N}} \phi,
\eeq
leading to an approximate $U(1)$ symmetry,
\beq
\phi \rightarrow e^{i \alpha} \phi.
\eeq
We will assume suppression of higher dimension operators by the scale $M_p$.  Suppose that the leading PQ symmetry-violating operator is:
\beq
\delta V = M_p^{4-N} \left (\gamma \phi^N + {\rm c.c.} \right ).
\label{eq:deltaV}
\eeq
Writing $\phi = f_a e^{ia/f_a} \equiv f_a e^{i \theta}$,
and assuming $\gamma \sim 1$, with an order one phase, the effective $\theta$
is roughly
\beq
\theta = {\rm Im}~ {\gamma} \frac{M_p^4 }{ f_\pi^2 m_\pi^2} \left(\frac{f_a }{ M_p} \right )^N.
\eeq
If, say, $f_a = 10^{11}$ GeV, then $\theta < 10^{-10}$ requires $N\ge 12$.  

Consider, instead, the requirement that the axion be the dark matter.  In \cite{cdf} a number of cosmological
scenarios were considered, resulting in different constraints on $N$.  But a minimal requirement is that the lifetime of the axion should be longer than the age
of the universe\footnote{Conceivably, the lifetime could be slightly shorter, being constrained only
by the requirement of structure formation.  Alternatively, effects of the radiation due to decaying axions could yield a stronger
constraint.  Observationally, the constraint is significantly stronger\cite{observational}. This range translates into about six orders
of magnitude in $m_a^2$, which is typically a change of $1$ or $2$ in the constraint on $N$, for a reasonable range of $f_a$.}.   Again with $f_a = 10^{11}$ GeV, the requirement is 
$m_a < 10^{-3} ~{\rm GeV}$, or $N>9$.  This is a weaker requirement on $N$ than the demands of $\theta$.
More stringent requirements can arise
from a detailed cosmological picture.  For example, things may be different if there is some approximate supersymmetry and the universe is dominated by a saxion
for some period.   Depending on the details, the requirement of dark matter can sometimes account for a small enough $\theta$.  It is also
possible, of course, that one has a bit of luck --- that $N$ is somewhat larger than it needs to be.

String theory suggests different possibilities\cite{cdf}.  
The PQ symmetry-violating operator in eqn.~\ref{eq:deltaV} may be suppressed by a small parameter, $A$, e.g $A = e^{-\frac{2\pi}{\alpha}}$, with $\alpha$ some small coupling constant.
%
%
$A$ might be extremely small; if the dark matter requirement is a suppression of the mass by $A^2$, for example, this
might also be sufficient to account for the quality of the Peccei-Quinn symmetry.

One concludes from this that it is {\it plausible} that anthropic considerations could favor an axion suitable to solve the strong CP problem, but that it is by no means certain; many cosmological and microphysical details would need to be understood to settle the question.

\section{Models Which Achieve a $\theta$ Discretuum}
\label{modelswhichwork}

The basic structure of the potential of Kaloper and Terning is rather puzzling.  In their picture, $\theta$ is continuous, and
its potential,
 for any choice of the fluxes, has a minimum
at $\theta =0$.  The QCD contribution to the potential dominates over other microscopic contributions.  It is not clear how $\theta$, as a continuous
variable subject to a superselection rule, scans. (KT refer to an earlier paper of Linde's\cite{lindetheta} which does not really provide a precise picture).  But if somehow $\theta$ is selected from this distribution, it must be compatible with the anthropic bound for the c.c.  Then, at least for some range of parameters, they argue,
small cosmological constant implies small $\theta$.

In this section, instead, we consider models which create a discretuum of values of $\theta$, and ask about the distribution of ground state energies with $\theta$.    
High energy dynamics give rise to a large number of degenerate vacua; the degeneracy is lifted
by QCD.

We start by revisiting the idea of the {\it irrational axion}\cite{irrationalaxion}, noting that this is, in fact,
a possible setting for these ideas.  However, it is not clear whether the irrational  axion is realized in any underlying
theory, so we then consider two other possible models.  These models are more concrete.  On the other hand,
while these theories generate a discretuum with a non-zero $\theta$,
to actually play a role in the cosmological constant problem, the discretuum must
be extremely fine, and the theories then exhibit some
rather implausible features.  For example, the model with the 
simplest field content requires an $SU(N)$ gauge group with $N > 10^{22}$, a number vastly larger than any appearing in proposed
landscape models.

\subsection{Prelude:  The Irrational Axion}

The irrational axion is a hypothetical setting with many vacua with different $\theta$\cite{irrationalaxion}.  If it were realized in an underlying theory, it might provide a setting for the ideas of \cite{kt}.   
In the irrational axion proposal, the $\theta$ potential receives contributions with different periodicities, which are not rational
multiples of one another.  This could arise from two groups, for example, with couplings to the same axion,
\beq
\sum_{i=1}^2\frac{a}{16 \pi^2 f_a} q_i F_i \tilde F_i,
\eeq
where $q_1$ is not a rational multiple of $q_2$.  

Then, taking the group $2$ to be the Standard Model $SU(3)$, and the group $1$ another group with scale $M^4
 \gg m_\pi^2 f_\pi^2$, 
 \beq
V(\theta) =- M^4 \cos (q_1 (\theta - \theta_0)) - m_\pi^2 f_\pi^2 \cos(q_2 \theta).
\label{irrationalpotential}
\eeq
For simplicity take $q_1 = x$, with $x$ irrational, and $q_2=1$.  Then the system has an infinity of nearly degenerate vacua with $\theta \approx \frac{2 \pi n }{x} + \theta_0$, hence
 a true discretuum of values of $\theta_{QCD}$.
In this case, $\Delta \theta$ is zero.

One has, then, a picture where for each value of the fluxes, there is a distribution of states of different $\theta$, with minimal energy at the point at which the strong interactions preserve CP.   The crucial element here is the absence of corrections to the potential of eqn.~\ref{irrationalpotential}, other than those from QCD, which lift the degeneracy.  Lacking a model, it is difficult to address the question of what sorts of corrections might arise to eqn.~\ref{irrationalpotential}.

One would have a similar picture if, say, $q_2/q_1$ were
not irrational, but
 a ratio of two extremely large primes.   In any case, as discussed in \cite{irrationalaxion} and subsequently
by others, it is not clear if such an axion actually emerges in string theory.
We will see shortly, however, that in this model, selection for the cosmological constant favors large, rather than small, $\theta$.

Ref.~\cite{bachlechner} does not directly address our questions here, but explores the interesting possibility that a theory with many axions might realize some features of the irrational axion.
In this situation, the number of states may be exponentially large.  One obtains bands of cosmological constant.  However, small c.c.~and small $\theta$ are not immediately correlated.   We will discuss a variant with many axions
in Section~\ref{sec:manyaxions}.

\subsection{Models with a single axion}
\label{singleaxion}

We consider in this section a more concrete model for the small $\Delta \theta$'s
required to tune the cosmological constant.  The model (if it is to yield extremely small $\Delta \theta$) is not particularly plausible, but illustrates
the main ingredients required to implement the KT solution.  We will see that for a limited range of parameters, one can
account simultaneously for both the observed c.c.~and the limits on $\theta$.  Conceivably there exists a more compelling
structure with the features we describe below.  We will offer another model in an appendix.

The model has two sectors, actual QCD and an $SU(N)$ gauge theory with a single adjoint fermion, $\lambda$ ($N$ will be {\it extremely} large), and a $Z_P$ symmetry acting on $\lambda$.  We include a complex scalar, $\phi$, coupled to
$\lambda$ and to a (heavy) quark:
\beq
\phi \lambda \lambda + \phi \bar Q Q.
\eeq
Under the $Z_P$ symmetry, $\phi \rightarrow e^{\frac{2 \pi i }{ P}} \phi$.  For general $P$, the discrete symmetry is anomalous with respect to both groups.  We will
comment on this in a moment.  We assume that $\phi$ develops a vev, breaking the approximate
Peccei-Quinn symmetry of the model,
\beq
\phi = f e^{i \theta}.
\eeq  Integrating out the heavy fields yields $F \tilde F$ couplings to both groups:
\beq
\frac{N \theta }{ 16 \pi^2} F^\prime \tilde F^\prime + \frac{\theta }{ 16 \pi^2} F \tilde F.
\eeq
This yields a potential
\beq \label{potentialN}
V =- \Lambda^4 \cos(N\theta) - m_\pi^2 f_\pi^2 \cos(\theta).
\eeq 

Both terms in \ref{potentialN}, for general $P$, violate the $Z_P$ symmetry. The phenomenon of
discrete symmetries with different apparent anomalies does occur in
string theory.  In such cases, these anomalies are cancelled by a Green-Schwarz mechanism involving multiple
axions.   In principle, these axions need not be light if their role is to cancel a discrete anomaly, but actual
realizations in string constructions involve light scalars.

The huge value of $N$ is puzzling from a string theory perspective.  Even focusing on the Cartan subalgebra of an $SU(N)$ group with an
exponentially large value of $N$, we are not aware of string theory constructions with such vast numbers of vector fields.
So these models do not appear extremely plausible.  Such vast numbers of fields would likely have other significant physics implications, which
we won't explore here.

Note here we do not have to assume an alignment of the CP conserving points in the two theories.   Rather we need to assume
that any additional contributions to the potential from other sources (such as other gauge groups) behaving
as, say, 
\beq
\delta V =\epsilon M^4 \cos(\theta- \theta_0),
\eeq
where $M$ indicates some fundamental scale and $\epsilon$ is a small parameter, are adequately suppressed.  Then we require $\epsilon M^4 < 10^{-10} m_\pi^2 f_\pi^2$.
This is similar to the requirement in theories with a light axion of a Peccei-Quinn symmetry of sufficient {\it quality} to solve the strong CP problem.  It imposes a minimum on $P$.

Neglecting effects of QCD, the minima of $\theta$ lie at points:
$\theta = \frac{2 \pi k }{ N}$.  So steps in $\theta$, are of size
\beq
\Delta \theta = \frac{2 \pi }{ N}.
\eeq 
In the absence of QCD, these states are degenerate. 
In the presence of QCD there is a contribution to the vacuum energy
\beq
 V = m_{\pi}^2 f_\pi^2 \left[ 1 - \cos(k \Delta\theta) \right] \, ,
\eeq
with $k$ an integer, which reduces to eqn.~\eqref{Vstep} for small $\Delta \theta$. To scan the c.c.~finely enough, the requirement $\Delta\theta < 10^{-22}$ here corresponds to $N > 10^{22}$.

In the appendix, we provide an alternative model, based on the {\it clockwork axion} idea\cite{Choi:2014rja, clockworkaxion}, which does not
require such huge gauge groups.

\subsection{Models with Multiple Axions}
\label{sec:manyaxions}

Taking our clue from the irrational axion idea, we can proceed in another direction.
If one is willing to pay the price of a large number of gauge groups (say 10), with large fermion representations (more generally with
large anomaly coefficient), one can avoid the gigantic single gauge group.  There are still stringent requirements regarding discrete symmetries.  The idea is to have, say, $M$ gauge groups (plus one more, QCD) and $M$ approximate PQ symmetries (so $M$ axions). The $M$ approximate discrete symmetries arise as a consequence of $M$ $Z_P$ types symmetries.  There are $M$ axions.   Take the fermions to be in the adjoint representation of the groups (or possibly larger representations --- the point is to have big anomalies, so large $\cos(N\theta)$ type terms).  More generally, one has something like
\beq
 V= \sum \Lambda_i^4 \cos(N_i (a_j \theta_j)) .
 \eeq
 This has a large number of degenerate solutions,
 \beq
 {\cal N} = \prod_{i=1}^{M} N_i ,
 \eeq
which can readily be huge.

 An example that is simple to analyze contains $M$ groups, $SU(N_i)$, $i=1,\dots ,M$; $M$ discrete (and approximate continuous) symmetries; and $M$ scalars.  Under the symmetries,
 \beq
 \phi_i \rightarrow e^{\frac{2 \pi i k }{ n_i}} \phi_i;~~\phi_i = f_i e^{i \theta_i};~~~\theta_i = \frac{a_i }{ f_i}.
 \eeq 
 Each scalar couples to adjoint fermions, $\phi_i \psi_i \psi_i$.  The resulting potential is
\beq
V =- \sum \Lambda_i^4 \cos(\theta_i N_i).
\eeq
The (degenerate) vacua have
\beq
\theta_i = \frac{2 \pi k }{ N_i}.
\eeq
The discrete symmetries are anomaly free if $n_i = p ~N_i$, with integer $p$ (with different conditions if the fermions are in representations
other than the adjoint representation).  Whether this is a condition we need to impose will be discussed later.

It is important that the vacua be degenerate to a high degree of approximation, with splittings smaller than $10^{-10} m_\pi^2 f_\pi^2$.
As it stands, the symmetries allow couplings
\beq
\delta {\cal L}= \frac{\gamma }{ M_p^{n_i -4}} \phi^{n_i} + {\rm c.c.}  .
\eeq
Depending on $f_i$, one obtains different conditions on $n_i$, but inevitably $n_i$ must be rather large --- typically $12$ or more.

Now suppose the QCD axion arises from a field $\phi_{QCD}$ with $Z_N$ charges $(q_1,\dots ,q_m)$, with couplings to a heavy quark $\phi \bar Q Q$. Also suppose
$\Lambda_{QCD} \ll \Lambda_i$.  Then the QCD contribution is a small perturbation, of the desired type, lifting the degeneracy among the vacua.
Note
\beq
\phi_{QCD}= e^{i \theta_{QCD}},
\eeq
where
\beq
\theta_{QCD} = \sum q_i \theta_i.
\eeq
This construction is also quite complicated.  An elaborate
set of fields and couplings will be required to generate a vev for $\phi_{QCD}$, and to avoid additional approximate global
symmetries.  But at least there is no group $SU$(Avogadro's number).  

\subsection{A Stringy Variant}

Rather than postulate a large set of gauge groups, we might consider a string theory with a large number of axions, and suppose that some
non-perturbative effect (e.g.~instantons in the string theory) gives rise to a potential for the axions.  Such a possibility has been considered
as an alternative to fluxes to obtain a large discretuum of states in  \cite{bachlechner}.  Here we are essentially considering both fluxes and
multiple axions as sources of the c.c.~(this will be quite explicit in the next few sections).  This potential might be a sum
of terms:
\beq
\sum_{a=1}^N \cos\left ( \sum_{i=1}^M \theta_i r_i^a \right) M_p^4 e^{-s_a},
\eeq
where $s_a$ might be the (assumed large) expectation value of some modulus, and $r_i^a$ are some integers.  We have rather arbitrarily
chosen $M_p$ as the fundamental scale.  The sum is over those terms which are large compared to QCD.  If $M=N$, there are a large (discrete) number of degenerate minima.   If $M<N$, there are not a large number of degenerate solutions (this is essentially the idea in \cite{bachlechner}
for generating a discretuum of cosmological constants).  If $M>N$, there are one or more light axions.

\section{Canceling the Cosmological Constant in Different Parameter Ranges}
\label{canceling}

In this section, we assume that the underlying theory has a discretuum of states of different $\theta$, and we ask whether anthropic selection for the
c.c.~leads to small $\theta$.
We start from eqn.~\eqref{cc1}, and we treat both $\theta$ and $\Lambda_{BP}$ as discrete parameters. 
We write $\theta = k \Delta\theta$, with $k$ an integer, and $\Lambda_{BP} = - q \Delta\Lambda$, with $q$ taking discrete
values, not necessarily integer, ranging from order 1 to large values, in order to stress that the spacing of states in the 
Bousso-Polchinski landscape is not uniform. 
The cosmological constant,
\beq \label{ccdiscrete}
\Lambda = -q  \Delta \Lambda -\cos (k \Delta \theta) m_\pi^2 f_\pi^2 \, \approx -q  \Delta \Lambda + \frac{1}{2} k^2 \Delta \theta^2 m_\pi^2 f_\pi^2 \, ,
\eeq
must satisfy the anthropic bound
\beq
0< \Lambda < \Lambda_a \, ,
\label{anthropicbound}
\eeq
with $\Lambda_a = 3 \times 10^{-47} \ {\rm GeV}^4$.
For given step sizes $\Delta \Lambda$ and $\Delta \theta$, there is a definite number of states [labeled
by $(k, q)$] that satisfy the bound of eqn.~\ref{anthropicbound}.
The question we wish to ask is:  are these states characterized by large or small $\theta$?

Given that the cosmological constant is a sum, in this picture, of two independent parameters, one might be skeptical
as to whether or how small $\theta$ might be favored.   Indeed, as we will explain in more detail below, if scanning is arbitrarily precise in
both $\Delta \Lambda$ and $\Delta \theta$, then small $\theta$ will not be favored.  It is perhaps slightly less obvious what happens if
scanning is coarse in both parameters.  The rest of this section enumerates and explores the various possibilities.

At a simple-minded level, there are a few constraints on the parameters that appear in eqn.~\eqref{ccdiscrete}. 
There is a maximum possible $k$, $k_{max}$, given by $k_{max} \Delta \theta = 2 \pi$ (there are also degeneracies in $k$, which will not
be particularly important in what follows).
Also, we require $\Delta \Lambda < m_\pi^2 f_\pi^2$, 
or we won't, in general, have any possibility of canceling off the c.c.~as we scan in $k$.

We organize the discussion according to whether we are scanning finely or coarsely in the $\Lambda_{BP}$ direction:
\begin{align}
{\rm Fine \ scanning}  & {} & \Delta\Lambda < \Lambda_a   \, , \\
{\rm Coarse \ scanning}  & {} & \Lambda_a < \Delta\Lambda < m_\pi^2 f_\pi^2     \, . 
\end{align}

\subsection{Fine $\Delta\Lambda$}

For $\Delta\Lambda < \Lambda_a$, we write $y = -q \Delta \Lambda$ and treat $y$ as a continuous parameter. 
Considering first $\Delta \theta \ll 10^{-22}$, we can approximate $\theta$ as continuous as well. 
The c.c.~is
\beq
\Lambda = y + \frac{1}{2} m_\pi^2 f_\pi^2 \theta^2.\label{eq:ccy}
\eeq
Because in this case we scan finely in $\Delta \Lambda$, it makes sense to define:
\beq
y_{max} = \Lambda_a - \frac{1}{2} m_\pi^2 f_\pi^2 \theta^2;~~~y_{min} = - \frac{1}{2} m_\pi^2 f_\pi^2 \theta^2.
\eeq 
We expect a uniform distribution in $y$ and $\theta$, so the fraction of states satisfying the anthropic condition is:
\beq
\int d \theta dy\  \Theta(y_{max} -y) \Theta(y-y_{min}) = \int d\theta \ (y_{max} - y_{min}),
\eeq
which is independent of $\theta$. So large $\theta$ is favored.

\begin{figure}
\begin{center}
\includegraphics[width=0.49\linewidth]{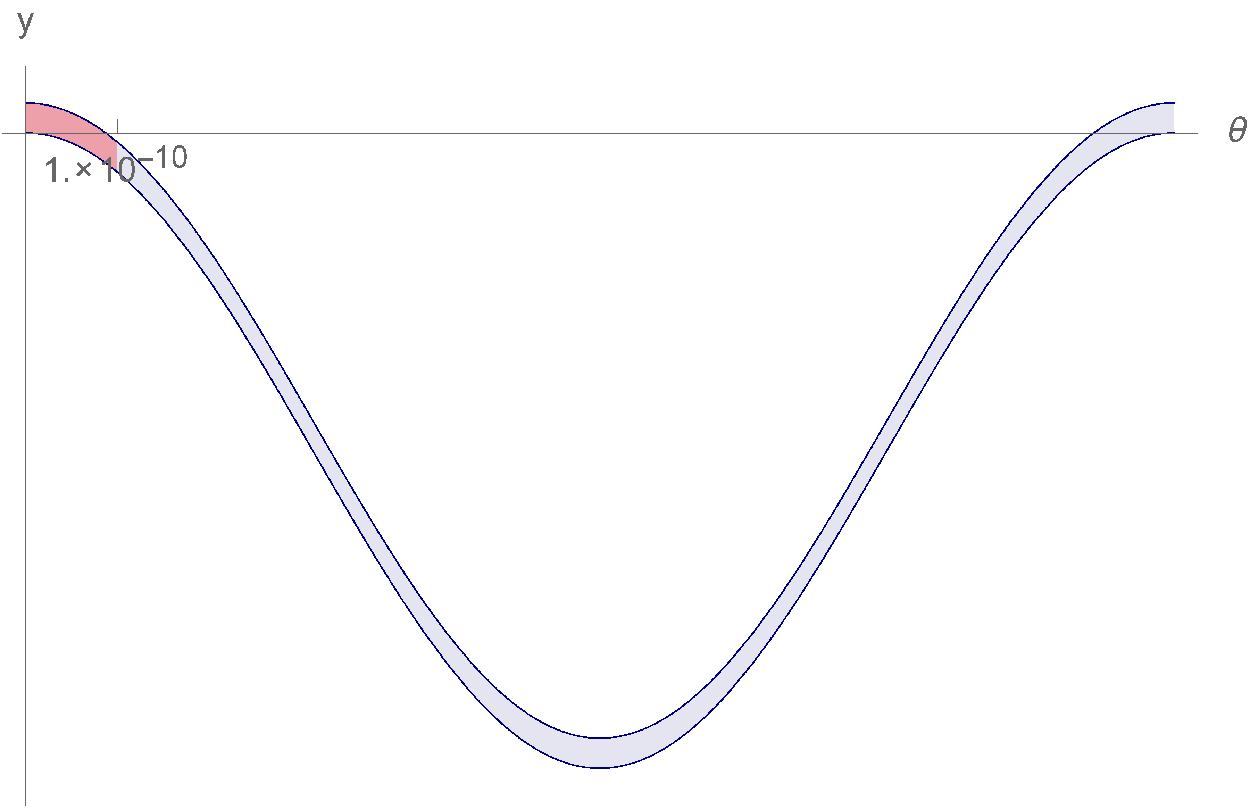}
\includegraphics[width=0.49\linewidth]{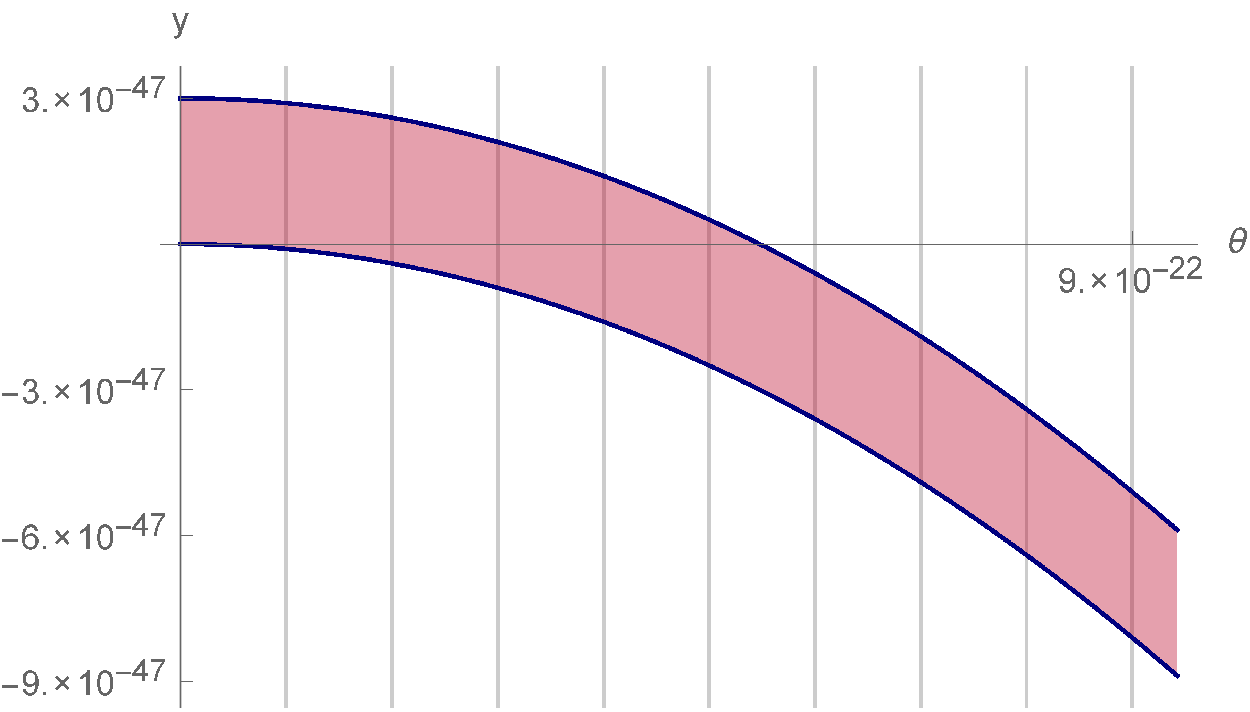}
\end{center}
\caption{Left: The shaded region is the portion of the region in $(\theta, y)$ space that produces an anthropically allowed value for the c.c., as determined by eqn.~\eqref{eq:ccy}. The red shading is the portion of this region for which $\theta<10^{-10}$ [not drawn to scale]. This shows that values of $\theta$ of order one are favored by the c.c.~anthropic selection. Right: zoomed in to very small $\theta$, drawn to scale.} \label{Fig:area}
\end{figure}

The same conclusion holds even if $\Delta\theta > 10^{-22}$ and we treat it as a discrete variable. Also in this case 
there is a larger number of states with large $\theta$, and the energy density of any such state can be canceled to the
desired accuracy by $y$ [see Fig.~\ref{Fig:area}].

\subsection{Coarse $\Delta\Lambda$}

In this subsection, we consider the case $\Lambda_a < \Delta \Lambda < m_\pi^2 f_\pi^2$.
Here, because we do not scan finely in $\Delta\Lambda$, we can't treat $q$ as a continuous variable as in our previous analysis. 
It is convenient to define
\be \label{abdef}
a \equiv \frac{m_\pi^2 f_\pi^2 \Delta\theta^2}{2 \Lambda_a} \, , \qquad b \equiv \frac{\Delta\Lambda}{\Lambda_a} \, , \qquad  1 < b < \frac{m_\pi^2 f_\pi^2}{\Lambda_a} \, .
\ee
As the theta potential
 goes as $k^2$, there can be a value, $k_0$, above which also the scanning in the theta direction becomes coarse (larger than $\Lambda_a$). 
$k_0$ is obtained from:
\be
\Delta V(k_0) = m_\pi^2 f_\pi^2 k_0 \Delta\theta^2 = \Lambda_a \, , \qquad \Rightarrow \quad k_0 = \frac{1}{2 a} \,
\ee 
We distinguish 2 cases. In the first
\be \label{smalla}
0< a < \frac{1}{2} \frac{\Lambda_a}{m_\pi^2 f_\pi^2} \, ,
\ee
we scan finely for all values of $k,~k\le k_{max}$. In the second
\be \label{largea}
\frac{1}{2} \frac{\Lambda_a}{m_\pi^2 f_\pi^2} < a < \frac{1}{2} \, ,
\ee
$1 < k_0 < k_{max}$, so the scanning becomes coarse for $k > k_0$.

Eq.~\eqref{anthropicbound} in terms of $a$ and $b$ becomes
\be \label{anthrok}
\frac{qb}{a} < k^2 < \frac{qb}{a} + \frac{1}{a} \, .
\ee
Now we want to ask: are there more
$(k,q)$
states satisfying the above inequalities at small or large
$k$?

We can write eq.~\eqref{anthrok} as
\be \label{anthropm}
k_- < k < k_+ \, ,
\ee
with
\be \label{kexp}
k_- =  \sqrt{\frac{qb}{a}} \, , \qquad  k_+ =  \sqrt{\frac{qb}{a} + \frac{1}{a}} \approx  \sqrt{\frac{qb}{a}} + \frac{1}{2 \sqrt{aqb}} \, .
\ee

Consider, first, the case of eqn. \ref{smalla}.  Then the number of states, for fixed $q$, satisfying the anthropic
constraint is
\be
N(q) \sim k_+ - k_- \approx  \frac{1}{2 \sqrt{aqb}} < 1 \, ,
\ee 
Treating $q$ and $\theta$ as approximately continuous, $q \sim \theta^2$, and we see that the number of states satisfying  the anthropic condition is larger at large $\theta$ ($q$) by $\sqrt{q} \propto \theta$.

Now take the case of eqn. \ref{largea}.
Let's start with fixing large $q$, and correspondingly large $\theta$, where, for general $q$ there is not a choice of $k$
satisfying the anthropic condition.  In other words, for 
\be \label{largeq}
\sqrt{\frac{qb}{a}} > k_0 \, \quad  \Rightarrow \quad \frac{1}{2\sqrt{aqb}} < 1 \,
\ee
the window $k_+ - k_-$ in this case is smaller than 1. Still, there is a small chance, for any given $q$, to find an integer
$k$ in such a window. The probability is
\be
P(q) \sim k_+ - k_- \approx  \frac{1}{2 \sqrt{aqb}} < 1 \, ,
\ee 
which decreases with increasing $q$ ($\theta$). 
The point, however, is that at large $q$ the number of possible states increases.
We can estimate the number of large $q$ states which satisfy the anthropic condition as
\be
N_{\rm large} = \int_{\frac{a}{b}k_0^2}^{\frac{a}{b}k_{max}^2} P(q) dq = \frac{1}{\sqrt{2a}b} \sqrt{\frac{m_\pi^2 f_\pi^2}{\Lambda_a}} - \frac{1}{2ab} \, .
\ee
Note that when $a$ saturates the upper bound of \eqref{smalla}, $N_{\rm large} = 0$, and remains zero for smaller $a$.

Next, consider small $q$:
\be \label{smallq}
\sqrt{\frac{qb}{a}} < k_0 \, \quad  \Rightarrow \quad \frac{1}{2\sqrt{aqb}} > 1 \, .
\ee
The expansion in \eqref{kexp} still holds, because $qb > 1$. Now, however, we have
\be
k_+ - k_- \approx  \frac{1}{2 \sqrt{aqb}} > 1 \, ,
\ee
implying there is at least one integer $k$ which satisfies \eqref{anthropm}, for any $q$ which satisfies \eqref{smallq}.
Thus, we can estimate the number of states which satisfy the anthropic condition in this case as 
\be
N_{\rm small} \approx \int_{1}^{\frac{a}{b}k_{0}^2} dq =  \frac{1}{4ab} - 1 \, .
\ee
Now we can compare $N_{\rm small}$ to $N_{\rm large}$.

In the first case of eq.~\eqref{smalla} we have $N_{\rm small} > N_{\rm large} = 0$, and the favored states have
$\theta < k_0 \Delta\theta$. However, $k_0 \Delta\theta > 1$ in this case, so large theta is favored.
In the second case of eq.~\eqref{largea}, $N_{\rm small}$ and $N_{\rm large}$ can be comparable as long as $a$ 
is close to the lower bound $\frac{1}{2} \frac{\Lambda_a}{m_\pi^2 f_\pi^2}$. This, again, favors $\theta \sim 1$. 
For larger values of $a$ we quickly obtain $N_{\rm large} \gg N_{\rm small}$, which also favors large theta.

When $a > \frac{1}{2}$ the chance of canceling the c.c.~is always smaller than one, and decreases
going to larger values of $q$. However, as in the analysis above, the bulk of the states which satisfy the anthropic condition
are at large $q$, so large
$\theta$ is favored.

\section{Conclusions}

Arguably if one cannot find a suitable solution to the strong CP problem in a landscape framework, the landscape idea may be 
unsupportable.  As a result, it is important to study any proposal to understand how the value of $\theta$ might be correlated with
other physical quantities which might be anthropically determined.  One possibility is that axions are selected by an anthropic
requirement for dark matter; the main issue is whether the Peccei-Quinn symmetry is of sufficient quality
to account for the small value of $\theta$.  We have reviewed the challenges to such a possibility, and concluded that such an
anthropic explanation of $\theta$ is plausible, but that whether it is realized depends on questions about the microphysical
theory and cosmology.

Kaloper and Terning propose to correlate $\theta$ with the
value of the cosmological constant.  
We have attempted to flesh out this proposal.  Rather than a continuous range of $\theta$, which, as we have
explained, is likely
to correspond to a conventional axion, we have argued that one should consider the possibility that, absent QCD,
there is a {\it massive} axion with a vast number of nearly degenerate ground states.  QCD then
lifts this degeneracy.  We have put forward models in which $\theta$ takes on a discretuum
of discrete values, with a potential on this discretuum of the desired type.  Having reduced the system to a discrete system,
we were able to assess the probability of finding larger or smaller $\theta$, in the sense of asking:  are most of the states
with anthropically favored c.c.~at large or small $\theta$?  The system has two parameters; 
on all of the space, $\theta \sim 1$ is favored.  A skeptical reader might have expected such a result from the start.  But it is
perhaps not completely obvious, at first glance, if scanning in $\Delta \theta$ and $\Delta \Lambda$ are not arbitrarily fine.
Here we have seen that, throughout the range of parameters, $\theta$ of order one is more likely.

It is conceivable that some other consideration might favor small $\theta$ within a $\theta$ discretuum.
The models we have 
proposed to obtain such a discretuum are not particularly attractive; indeed they are hardly plausible.  
The irrational axion\cite{irrationalaxion} has some of the desired features
of such a system, but it is not clear that such axions actually arise in any theory of quantum gravity, and, in any
case, this would predict $\theta \sim 1$.
Another class of models involves a huge gauge group and a large discrete symmetry; others a very large
number of fields.

As for the anthropic axion, in addition to the question of axion quality, mentioned above, it should be stressed that existing anthropic arguments
for the dark matter density are interesting but arguably not compelling.
While these are serious concerns, given the challenges of tying small $\theta$ to the cosmological constant, an anthropic axion 
would seem a more plausible
solution of the strong CP problem in this framework.  The reader, of course, is free to view all of this as reason for skepticism
about the landscape program altogether.

\vskip 1cm
\noindent
{\bf Acknowledgements:}  This work was supported in part by the U.S. Department of Energy grant number DE-FG02-04ER41286. 
L.U. acknowledges support from the PRIN project ``Search for the Fundamental Laws and Constituents'' (2015P5SBHT\textunderscore 002).
 L.S.H. is supported by the Israel
Science Foundation under grant no.~1112/17.  We thank Patrick Draper for extensive conversations.  M.D. also thanks Savas Dimopoulos and Peter Graham for a helpful conversation.  We especially thank Clifford Cheung and Prashant Saraswat for correcting a misconception which seemed to allow an anthropic solution for a narrow range of $\Delta \theta$ and $\Lambda$.

\appendix

\section{A Clockwork Construction}

The construction relies on the potential~\cite{clockworkaxion}
\begin{equation} \label{clockpot}
V(\Phi) = \sum_{j=1}^{N+1} \left( -\mu^2_\Phi \Phi_j^\dagger \Phi_j + \frac{\lambda_\Phi}{4} | \Phi_j^\dagger \Phi_j |^2 \right)
+ \sum_{j=1}^{N} \left( \epsilon_\Phi \Phi_j^\dagger \Phi^3_{j+1} + {\rm h.c.}  \right) \, ,
\end{equation}
where $\Phi_j$'s are complex scalar fields.
The terms in the first sum respect a global $U(1)^{N+1}$ symmetry, while the second sum explicitly breaks it to a $U(1)$. 
The fields $\Phi_j$ have charges $Q = 1, \frac{1}{3}, \frac{1}{9}, \dots, \frac{1}{3^N}$ under the unbroken $U(1)$.
We take $\mu^2_\Phi > 0$, so all the $U(1)$'s are spontaneously broken at a scale $f = \sqrt{(2\mu^2_\Phi)/\lambda_\Phi}$. 
All the radial modes then have a mass of order $f$. Neglecting the second sum in the potential, we have $N+1$ 
massless Nambu-Goldstone bosons (NGBs). Taking into account the second sum, with the explicit breaking parameter 
$\epsilon_\Phi \ll 1$, we find that $N$ of these NGBs get a mass
 of order  $\sqrt{\epsilon_\Phi} f$, while one, $\phi$, remains massless. The latter corresponds to the linear combination 
 $\phi = {\cal N} \left( 1 \ \frac{1}{3} \ \frac{1}{9} \ \dots \frac{1}{3^N} \right)$. 
 Here ${\cal N}$ is a normalization factor. 
 
Now, we can write the Yukawa coupling $\Phi_{N+1} \bar \psi_{N+1} \psi_{N+1}$, where $\psi_{N+1}$ is a fermion that lives
at the site $N+1$ and is in the fundamental representation of $SU(3)_c$, the QCD gauge group. This is a KSVZ axion model,
and the QCD anomaly leads to the coupling
\begin{equation}
\frac{\alpha_s}{8\pi} \frac{\phi}{F} G \tilde G \, ,
\end{equation} 
which in turn gives us the cosine potential
\begin{equation}
V_{\rm QCD} = m_\pi^2 f_\pi^2 \left( 1 - \cos \frac{\phi}{F} \right) = m_\pi^2 f_\pi^2 \left( 1 - \cos \theta \right) \, .
\end{equation}
Here we have defined $\theta \equiv \frac{\phi}{F}$, and we have $F = 3^N f$ from the clockwork construction.

Next, we introduce fermions $\psi_1$ in the fundamental of a new gauge group $SU(N_1)$, with confinement scale 
$\Lambda$, and we write the Yukawa coupling $\Phi_1 \bar \psi_1 \psi_1$. 
Again there is an anomaly with respect to $SU(N_1)$, from which we get the potential
\begin{equation}
V_1 = \Lambda^4 \left( 1 - \cos \frac{\phi}{f} \right) = \Lambda^4 \left( 1 - \cos (3^N \theta) \right) \, .
\end{equation}
Note that, as this potential arises from a coupling at the first clockwork site where the axion $\phi$ is mostly localized, 
the periodicity is given by $f$.
 
The sum $V_{\rm QCD} + V_1$ gives the desired structure of eqn.~\eqref{potentialN}.

\bibliography{anthropic_theta}{}
\bibliographystyle{utphys}

\end{document}